\documentclass[reprint,twocolumn,superscriptaddress,amsmath,amssymb,nofootinbib,preprintnumbers]{revtex4-2} 

\usepackage{mathptmx}
\usepackage{graphicx}  
\usepackage[mathlines]{lineno}
\usepackage[english]{babel} 
\usepackage{xcolor}
\usepackage{ulem}
\usepackage{dcolumn}   
\usepackage{bm}        
\usepackage{amssymb}   
\usepackage{amsbsy}   
\usepackage{comment}
\usepackage{multirow}
\usepackage{diagbox}
\usepackage[draft]{todonotes}
\usepackage{mathtools}
\usepackage{newtxtext}\usepackage[varvw]{newtxmath}
\usepackage{nicefrac}
\usepackage{tikz-feynman}
\tikzfeynmanset{compat=1.1.0}

\usepackage{hyperref}
\hypersetup{
    colorlinks=true,
    linkcolor=cyan,
    filecolor=magenta,      
    urlcolor=cyan,
    citecolor=violet,
}
\usepackage{cleveref}

\makeatletter
\g@addto@macro\bfseries{\boldmath}
\makeatother

\def \Hinf {H_\text{inf}}
\def \Mp {M_\text{P}}
\def \nn {\mathbf{n}}
\def \dif {\mathrm{d}}
\newcommand{\LL}{_\mathrm{L}}

\newcommand{\cc}{^\mathrm{c}}

\begin{document}

\title{Constraints on metastable superheavy dark matter coupled to sterile neutrinos with the Pierre Auger Observatory}

\author{A.~Abdul Halim}
\affiliation{University of Adelaide, Adelaide, S.A., Australia}

\author{P.~Abreu}
\affiliation{Laborat\'orio de Instrumenta\c{c}\~ao e F\'\i{}sica Experimental de Part\'\i{}culas -- LIP and Instituto Superior T\'ecnico -- IST, Universidade de Lisboa -- UL, Lisboa, Portugal}

\author{M.~Aglietta}
\affiliation{Osservatorio Astrofisico di Torino (INAF), Torino, Italy}
\affiliation{INFN, Sezione di Torino, Torino, Italy}

\author{I.~Allekotte}
\affiliation{Centro At\'omico Bariloche and Instituto Balseiro (CNEA-UNCuyo-CONICET), San Carlos de Bariloche, Argentina}

\author{K.~Almeida Cheminant}
\affiliation{Institute of Nuclear Physics PAN, Krakow, Poland}

\author{A.~Almela}
\affiliation{Instituto de Tecnolog\'\i{}as en Detecci\'on y Astropart\'\i{}culas (CNEA, CONICET, UNSAM), Buenos Aires, Argentina}
\affiliation{Universidad Tecnol\'ogica Nacional -- Facultad Regional Buenos Aires, Buenos Aires, Argentina}

\author{R.~Aloisio}
\affiliation{Gran Sasso Science Institute, L'Aquila, Italy}
\affiliation{INFN Laboratori Nazionali del Gran Sasso, Assergi (L'Aquila), Italy}

\author{J.~Alvarez-Mu\~niz}
\affiliation{Instituto Galego de F\'\i{}sica de Altas Enerx\'\i{}as (IGFAE), Universidade de Santiago de Compostela, Santiago de Compostela, Spain}

\author{J.~Ammerman Yebra}
\affiliation{Instituto Galego de F\'\i{}sica de Altas Enerx\'\i{}as (IGFAE), Universidade de Santiago de Compostela, Santiago de Compostela, Spain}

\author{G.A.~Anastasi}
\affiliation{Universit\`a di Catania, Dipartimento di Fisica e Astronomia ``Ettore Majorana``, Catania, Italy}
\affiliation{INFN, Sezione di Catania, Catania, Italy}

% ...

\collaboration{The Pierre Auger Collaboration}
\email{spokespersons@auger.org}
\homepage{http://www.auger.org}
\noaffiliation


\date{\today}

\begin{abstract}
 
\noindent Dark matter particles could be superheavy, provided their lifetime is much longer than the age of the universe. Using the sensitivity of the Pierre Auger Observatory to ultra-high energy neutrinos and photons, we constrain a specific extension of the Standard Model of particle physics that meets the lifetime requirement for a superheavy particle by coupling it to a sector of ultra-light sterile neutrinos. Our results show that, for a typical dark coupling constant of 0.1, the mixing angle $\theta_m$ between active and sterile neutrinos must satisfy, roughly,  $\theta_m \lesssim 1.5\times 10^{-6}(M_X/10^9~\mathrm{GeV})^{-2}$ for a mass $M_X$ of the dark-matter particle between $10^8$ and $10^{11}~$GeV.

\end{abstract}

\maketitle


Despite its countless successes, the Standard Model (SM) of particle physics is known to be incomplete. Missing elements include, among others, dark matter (DM) and neutrino masses. A minimal extension to the SM is to add right-handed neutrinos inert to SM interactions (``sterile'' neutrinos) that mix with left-handed ones active under weak interactions (``active'' neutrinos) via the Brout--Englert--Higgs mechanism. Sub-eV mass eigenvalues result from mixing angles made sufficiently small by highly-enough massive sterile neutrinos (seesaw mechanism~\cite{Gell-Mann:1979vob,Yanagida1979}). Other extensions may use additional sterile neutrinos, as their number is in general largely under-constrained. This is the case of the extension presented in~\cite{Dudas:2020sbq}, considered in this study as a beyond-standard-model (BSM) benchmark. In this model, which includes DM and neutrino masses, a second sterile neutrino is assumed to render metastable a superheavy particle referred to as $X$ with mass $M_X$ and lifetime $\tau_X$, the decay products of which can leave unique signatures in the data of the Pierre Auger Observatory~\cite{PierreAuger:2015eyc}.

A superheavy particle that is metastable attracts particular attention if it is to be a viable DM candidate. The constraints imposed on $\tau_X$ are indeed demanding: $\tau_X\gtrsim 10^{22}~$yr, see e.g.~\cite{Kachelriess:2018rty,Alcantara:2019sco,Ishiwata:2019aet,Guepin:2021ljb,Berat:2022iea,IceCube:2022clp,Das:2023wtk} for recent estimates in various channels. Compliance with these constraints indeed calls for adjusting a reduced coupling constant $\alpha_X$ down to a tiny level and, simultaneously, the multiplicity of the final state to a large value~\cite{deVega:2003hh,PierreAugerCollaboration:2022wir}. This is in general challenging for theoretical constructions, unless the decay rules are based on non-perturbative effects~\cite{Kuzmin:1997jua,PierreAugerCollaboration:2022tlw}. In the BSM benchmark~\cite{Dudas:2020sbq}, however, the lifetime does not resort to such a fine tuning. The DM particle $X$ interacts only with sub-eV and superheavy ($10^{12-14}~$GeV) sterile neutrinos, of masses $m_N$ and $M_N$ respectively, via Yukawa couplings $y_m$ and $y_M$. In the mass-eigenstate basis, neutrinos are then quasi-active or quasi-sterile, depending on the mixture of active and sterile neutrinos governed by a small mixing angle $\theta_m\simeq \sqrt{2} y_m\varv/m_\nu$, with $\varv$ the electroweak scale and $m_\nu$ the mass of the known neutrinos. To leading order in $y_m$, quasi-active neutrinos are produced from quasi-sterile ones subsequent to the decay of $X$. Consequently, the coupling $y_m$ controls the dominant decay channels and allows for trading a factor $(M_X/\Mp)^2$ (with $\Mp$ the Planck mass) for a $(m_\nu\theta_m/v)^2$ one in the decay width of $X$. This trading enables the reduction of the width by a factor $\sim 10^{-25}\theta_m^2$ for a benchmark value $M_X=10^9~$GeV, leading to the required lifetimes well beyond the age of the universe. 

Ultra-high energy (UHE) neutrinos and photons are expected to emerge from the cascade of the $X$ decay. In this Letter, we use the sensitivity of the Pierre Auger Observatory to such neutrinos and photons to impose constraints on the active-sterile neutrino mixing angle $\theta_m$ for DM masses $M_X\gtrsim 10^8~$GeV. After recalling the main features of the Observatory that enable the detection of neutrinos and photons, we present the main decay channels of the $X$ particle. We then calculate the number of  neutrinos and photons expected to be observed at the Observatory as a function of $\theta_m$ and $M_X$. Their non-observation allows us to constrain $\theta_m$ for $M_X\gtrsim 10^8~$GeV. In a last section, we comment on the complementary bounds obtained on $M_X$ from DM abundance and on $\theta_m$ from cosmological observations. \\

\textit{The Pierre Auger Observatory} is the largest ground-based observatory that exploits extensive air showers to study UHE cosmic rays~\cite{PierreAuger:2015eyc}. Several detection techniques are combined. A surface of 3,000~km$^2$ is covered with a ground array of particle detectors separated by 1,500~m surrounded by fluorescence detectors spread over four sites. This baseline configuration is complemented with low-energy enhancements: a first nested array of 24~km$^2$ with particle detectors separated by 750~m and overlooked by three additional fluorescence telescopes with an elevated field of view, and a second one of 1.14~km$^2$ with detectors separated by 433~m, some of them on top of buried scintillation modules used to measure the number of high-energy muons. The fluorescence detectors provide, during moonless nights, direct observation of the longitudinal shower profile, which allows for the measurement of the energy and the primary-mass sensitive depth of the shower maximum, $X_{\mathrm{max}}$. On the other hand, the ground-level and underground detectors sample the shower particles with a permanent duty cycle. Although showers are observed at a fixed slice in depth with this technique, their longitudinal development is embedded in the signals detected. 

Neutrinos of all flavors can be distinguished from nuclei through showers developing deeply in the atmosphere at large zenith angles (down-going detection mode)~\cite{Capelle:1998zz}, while tau neutrinos can undergo charged-current interactions and produce a $\tau$ lepton in the crust of the Earth that eventually decays in the atmosphere, inducing an upward-going shower (Earth-skimming detection mode)~\cite{Bertou:2001vm}. In both cases, neutrinos can be identified at the Observatory with signals in the ground-level particle detectors that are spread over time, by contrast to narrow ones expected from the much more abundant nuclei-induced showers at large zenith angles~\cite{PierreAuger:2007vvh,PierreAuger:2011cpc,PierreAuger:2015ihf}. Photon-induced showers have also distinct salient features~\cite{Billoir:2000uy,Risse:2007sd}. Their first interactions are of electromagnetic nature, and the transfer of energy to the hadron/muon channel is reduced with respect to the bulk of hadron-induced showers. This results in a lower number of secondary muons. Additionally, as the development of photon showers is delayed by the typically small multiplicity of electromagnetic interactions, their $X_{\mathrm{max}}$ is deeper in the atmosphere than for showers initiated by hadrons. 

Based on these characteristics, a number of analyses have been designed to make the most of the sensitivity of the different detectors of the Observatory to neutrinos above $\simeq 10^8~$GeV~\cite{PierreAuger:2019ens} and to photons above $5\times 10^7~$GeV~\cite{PierreAuger:2022uwd,Savina:2021cva,PierreAuger:2022aty,PierreAuger:2023nkh}. The non-observation of point sources and diffuse fluxes allowed the derivation of upper bounds that constrain various models very effectively. In the following, we draw constraints on the BSM benchmark~\cite{Dudas:2020sbq} from these non-observations. \\

\textit{In the BSM benchmark~\cite{Dudas:2020sbq},} the DM candidate is a pseudo-scalar particle $X$ coupled to the sterile neutrinos sector alone. The interaction can be represented diagrammatically as
\begin{equation}
\begin{tikzpicture} \begin{feynman}
   \vertex(a) {\(X\)};
   \vertex[right=of a] (b);
   \vertex[above right=of b] (f1){\(N\)};
   \vertex[below right=of b] (f3){\(N\)};
   \diagram* {
   (a)-- [scalar,edge label'=\(q_{\mu}\)] (b)-- [anti fermion] (f1),
   (b)-- [fermion] (f3),
   }; 
   \node at (3.2,0.) {\(-i\alpha_{X}q_\mu\gamma^\mu\gamma^5/M_{P},\)};
\end{feynman}\end{tikzpicture}
\label{eqn:2body}
\end{equation}
with $q$ the four-momentum of $X$, and $\gamma$ the Dirac matrices. Sterile neutrinos, which are right-handed, are denoted as $N$. Two types of Majorana sterile neutrinos are actually introduced. The first type, $N_M$, corresponds to the right-handed neutrino of the seesaw mechanism, with a superheavy mass $M_N$. The second type, $N_m$, with mass $m_N$, is necessary to allow the $X$ particle to decay in an electroweak channel. In the basis of interaction eigenstates, the couplings between sterile neutrinos, active neutrinos and SM Higgs isospinor scalar fields through Yukawa parameters $y_m$ and $y_M$ give rise, after symmetry breakdown, to Dirac masses $m^\mathrm{D}_{N_m}=y_m\varv/\sqrt{2}$ and $m^\mathrm{D}_{N_M}=y_M\varv/\sqrt{2}$. Assuming the hierarchy $m_N<m^\mathrm{D}_{N_M} \ll M_N$, the three mass eigenstates associated to the eigenvalues $(m_1,m_2,m_3)$ are well approximated by the following mixing between $\nu\LL$ (active neutrinos), $N_M$ and $N_m$ (and their conjugate partners with superscripts c):
\begin{eqnarray} 
\label{eqn:masseigenstates-stable1}
\nu_1 &\simeq & (N_m+{N_m}\cc)+\theta_m(\nu\LL+{\nu\LL}\cc), \\
\label{eqn:masseigenstates-stable2}
\nu_2 &\simeq & (\nu\LL+{\nu\LL}\cc)-\theta_m(N_m+{N_m}\cc), \\
\label{eqn:masseigenstates-stable3}
\nu_3 &\simeq & N_M,
\end{eqnarray}
with the mixing angle $\theta_m$, assumed to be $\ll 1$, defined as \mbox{$\theta_m\simeq y_m\varv/\sqrt{2}(m_1+m_2)$}. For small mixings, the first (second) mass eigenstate $\nu_1$ ($\nu_2$ ) is almost the light sterile (active) neutrino with a small admixture of the active (sterile) one, and that the third mass eigenstate $\nu_3$ is almost the superheavy sterile neutrino. Note that to avoid complications that would be irrelevant to the question of DM, the flavor couplings in the active neutrino sector are considered diagonal in this study. The three mass eigenvalues read as
\begin{eqnarray} \label{eqn:masseigenvalues-stable}
m_1 &\simeq & m_N, \\
m_2 &\simeq & y_m^2\varv^2/2M_N, \\
m_3 &\simeq & M_N.
\end{eqnarray}
The third eigenstate $\nu_3$ appears to decouple from the two other states. While the superheavy particle $N_M$ is essential to provide the mass to the active neutrino through a mixing angle $\theta_M=y_M\varv/2M_N$ identical to that of the standard seesaw mechanism~\cite{Gell-Mann:1979vob,Yanagida1979}, we shall ignore $\nu_3$ hereafter. From the constraints on SM neutrino masses $\sum m_\nu\leq 0.12~$eV inferred from cosmological observations~\cite{Planck:2018vyg}, we use hereafter $m_2\simeq 0.04~$eV as a benchmark. As for $m_1$, following~\cite{Dudas:2020sbq}, we use $m_1=10^{-4}~$eV to fix the ideas; we shall see below that as long as $m_1\ll m_2$, results are not sensitive to the specific choice. That $m_1 \ll m_2$ is required for the lifetime of $X$ to be much larger than the age of the universe.

To leading order in $\theta_m$, the interaction described by Eqn.~\ref{eqn:2body} gives rise, in the basis of mass eigenstates, to the two-body decay $X\rightarrow \nu_1\nu_1$. However, in the relevant parameter space such that $m_1 \ll m_2\ll m_e \ll m_Z \ll M_X$, the total width $\Gamma^X$ is dominated by three-body channels stemming from the diagram depicted in Eqn.~\ref{eqn:2body} and the interaction between active/sterile neutrinos and the Higgs isodoublet with Yukawa coupling $y_m\simeq \sqrt{2}\theta_mm_2/v$. The channel $X\rightarrow h\nu_1\nu_2$, diagrammatically represented as
\begin{equation}
\begin{tikzpicture} \begin{feynman}
   \vertex(a) {\(X\)};
   \vertex[right=of a] (b);
   \vertex[above right=of b] (f1){\(\nu_{1}\)};
   \vertex[below right=of b] (c);
   \vertex[above right=of c] (f2){\(h\)};
   \vertex[below right=of c] (f3){\(\nu_{2}\)};
   \diagram* {
    (a)-- [scalar] (b)-- [anti fermion] (f1),
    (b)-- [fermion,edge label'=\(\nu_{1}\)] (c),
    (c)-- [scalar] (f2),
    (c)-- [fermion] (f3),
   };
   \node at (3.6,-1.1) {\(\theta_m m_2/v,\)};
\end{feynman}\end{tikzpicture}
\label{eqn:3body-Xhnunu}
\end{equation}
gives the most important contribution to the width~\cite{Dudas:2020sbq}:
\begin{equation}
\Gamma_{h\nu_1\nu_2}^X=\frac{\alpha_X^2\theta_m^2}{192\pi^3}\left(\frac{M_X}{\Mp}\right)^2\left(\frac{m_2}{v}\right)^2 M_X. 
\label{eqn:width-hnunu}
\end{equation}
Due to the structure of the interaction depicted in Eqn.~\ref{eqn:2body}, a factor $(M_X/\Mp)^2$ expected from dimensional arguments~\cite{deVega:2003hh} is traded for a factor $(m_2/v)^2$.

Although subdominant, there are two other three-body decay channels of interest, namely $X\rightarrow Z\nu_1\nu_2$ and $X\rightarrow W\nu_1e$ with respective widths
\begin{eqnarray}
   \label{eqn:width-Znunu}
   \Gamma_{Z\nu_1\nu_2}^X&=&\frac{\alpha_X^2g^2\theta_m^2}{798\pi^3\cos^2{\theta_{\mathrm{W}}}}\left(\frac{M_X}{\Mp}\right)^2\left(\frac{m_2}{m_Z}\right)^2 M_X, \\
   \label{eqn:width-Wnue}
   \Gamma_{W\nu_1 e}^X&=&\frac{\alpha_X^2g^2\theta_m^2}{768\pi^3}\left(\frac{M_X}{\Mp}\right)^2\left(\frac{m_1}{m_W}\right)^2 M_X,
\end{eqnarray}
with $g$ the weak-isospin gauge constant and $\theta_{\mathrm{W}}$ the Weinberg angle. These two widths share the same structure as $\Gamma^X_{h\nu_1\nu_2}$ modulo smaller numerical factors. \\

\textit{UHE neutrinos and photons} are expected as byproducts of the decay of any particle with a mass much larger than the electroweak scale. Ultimately, they result from splitting-particle effects due to soft or collinear (real) radiative corrections enhanced by large logarithmic factors at high scale. The probability that a particle $a$ at a scale $\mu_a$ fragments to produce a particle $b$ at a scale $\mu_b$ carrying a fraction $x$ of the initial energy is described by a fragmentation function (FF) $D_a^b(x; \mu_a,\mu_b)$. The FFs are evolved starting from measurements at the electroweak scale up to the energy scale $M_X$. In the QCD sector, the evolution is governed by DGLAP equations to account for the splitting function that describes the emission of parton $k$ by parton $j$. The resulting prompt spectra of photons and neutrinos have been derived in~\cite{Aloisio:2003xj} and in several subsequent studies. Similarly, electroweak cascading can be described by evolution equations valid for a spontaneously broken theory~\cite{Ciafaloni:2001mu}. Seminal works of~\cite{Sarkar:2001se,Barbot:2002gt,Aloisio:2003xj} in the QCD sector and of~\cite{Berezinsky:2002hq} in the electroweak one have provided the calculation of the FFs to derive the prompt flux of high-energy secondaries from the decay of a particle at high scale. We use hereafter the up-to-date HDMSpectra tool~\cite{Bauer:2020jay} to calculate the energy spectra of UHE neutrinos and photons from the decay of the $X$ particle in the $X\rightarrow h\nu_1\nu_2$, $X\rightarrow Z\nu_1\nu_2$ and $X\rightarrow W\nu_1 e$ channels. 

The number of neutrinos $n_\nu(E)$ expected to be observed above an energy threshold $E$ results from the integration over the sky of the directional exposure $\mathcal{E}_\nu(E,\nn)$ of the Observatory and of the flux of neutrinos emitted isotropically in proportion to the DM density accumulated in galaxy halos~\cite{Aloisio:2015lva}:
\begin{multline}
\label{eqn:n_nu}
   n_\nu(E)=
   \iint_{\geq E} \dif\nn~\dif E' \frac{\mathcal{E}_\nu(E',\nn)}{4\pi M_X} \Bigg[\frac{\dif \Gamma_\nu(E')}{\dif E'} \int \dif s~\rho(\mathbf{x}_\odot+s\nn)\\
   +\Omega_X\rho_{\mathrm{c}}\int\dif z~\frac{e^{-S_\nu(E',z)}}{(1+z)H(z)}\frac{\dif \Gamma_\nu(E^\star)}{\dif E'}\bigg\rvert_{E^\star=(1+z)E'}\Bigg].
\end{multline}
The first contribution is from the Milky Way halo, in which the DM energy density is parameterized by a profile function $\rho(\mathbf{x})$. $\mathbf{x}_\odot$ is the position of the Solar system in the Galaxy, $s$ is the distance from $\mathbf{x}_\odot$ to the emission point, and $\nn\equiv\nn(\ell,b)$ is a unit vector on the sphere pointing to the longitude $\ell$ and latitude $b$, in Galactic coordinates. There are uncertainties in the determination of the galactic-halo profile. We use here the traditional NFW profile as a reference~\cite{Navarro:1995iw}, 
\begin{equation}
   \label{eqn:NFW}
   \rho_\text{DM}(R)=\frac{\rho_s}{(R/R_s)(1+R/R_s)^2},
\end{equation}
where $R$ is the distance to the Galactic center, $R_s=24$\,kpc, and $\rho_s$ is fixed by the DM density in the solar neighborhood, namely $\rho_\odot=0.44$\,GeV\,cm$^{-3}$~\cite{Jiao:2023aci}. To quantify the systematics stemming from the uncertainties on this profile, we repeat the analysis using other ones~\cite{Einasto:1965czb,Burkert:1995yz,Moore:1999nt}. The second contribution in Eqn.~\ref{eqn:n_nu}, which is obtained by integration over redshift $z$ and which amounts to about 10\% of the first one, is from all other galaxies. $\rho_\mathrm{c}$ is the critical energy density, $\Omega_X$ is the DM abundance, $S_\nu(E,z)$ is the neutrino opacity of the universe as calculated in~\cite{Gondolo:1991rn}, and the Hubble rate $H(z)$ depends on that observed today, $H_0$, and on the total matter abundance, $\Omega_\mathrm{m}$, through $H(z)=H_0\sqrt{\Omega_\mathrm{m}(1+z)^3+(1-\Omega_\mathrm{m})}$. In both contributions, the particle $X$ decays into a (SM) daughter particle $I$ whose FF leads to neutrinos. Thus, the differential decay width into neutrinos, $\dif\Gamma_\nu/\dif E$, results from the convolution of the differential decay width of $X$ to $I$ with the FF of $I$ into neutrinos. For a single flavor, it reads as
\begin{equation}
\label{eqn:dGammadE}
   \frac{\dif \Gamma_\nu}{\dif E}=\frac{2}{\Gamma^X}\sum_I\int_x^1\frac{\dif y}{y}\frac{\dif\Gamma^X_{I}(y)}{\dif y}D_I^\nu\left(\tfrac{x}{y},\tfrac{M_X}{2},0\right),
\end{equation}
with $x=2E_\nu/M_X$. In the $h\nu_1\nu_2$ channel, the differential decay width reads as
\begin{equation}    
   \label{eqn:diffrate-nu}
   \frac{\dif\Gamma^{X\rightarrow h\nu_1\nu_2}_\nu(y)}{\dif y} =3\Gamma^X_{h\nu_1\nu_2} y^2
\end{equation}
for neutrino final states, while it reads as 
\begin{equation}    
   \label{eqn:diffrate-h}
   \frac{\dif\Gamma^{X\rightarrow h\nu_1\nu_2}_h(y)}{\dif y} =6\Gamma^X_{h\nu_1\nu_2} y(1-y)
\end{equation}
for Higgs final states~\cite{Dudas:2020sbq}. Similar expressions hold in the cases of the $Z\nu_1\nu_2$ and $W\nu_1e$ channels, with the corresponding decay widths. Finally, the differential widths entering into Eqn.~\ref{eqn:dGammadE} account for all detectable flavors for the $\nu_{2\alpha}$ species, where an explicit flavor index $\alpha$ is re-introduced. In the down-going detection mode, the three flavors contribute explicitly as
\begin{multline}
   \frac{\dif \Gamma_{\nu_2,\mathrm{all}}}{\dif E}=\frac{2}{\Gamma^X}\int_x^1\frac{\dif y}{y}\Bigg[\sum_\alpha\Bigg(\frac{\dif\Gamma^{X\rightarrow h}_h(y)}{\dif y}D_h^{\nu_{2\alpha}}\left(\tfrac{x}{y}\right)  \\
   + \frac{\dif\Gamma^{X\rightarrow Z}_Z(y)}{\dif y}D_Z^{\nu_{2\alpha}}\left(\tfrac{x}{y}\right) 
   +2\frac{\dif\Gamma^{X\rightarrow W}_W(y)}{\dif y}D_W^{\nu_{2\alpha}}\left(\tfrac{x}{y}\right) \Bigg) \\
   +\sum_{\alpha,\beta}\Bigg(\frac{\dif\Gamma^{X\rightarrow h}_{\nu}(y)}{\dif y}D_{\nu_{2\beta}}^{\nu_{2\alpha}}\left(\tfrac{x}{y}\right) 
   +\frac{\dif\Gamma^{X\rightarrow Z}_{\nu}(y)}{\dif y}D_{\nu_{2\beta}}^{\nu_{2\alpha}}\left(\tfrac{x}{y}\right) \Bigg)\\
   +\sum_\alpha 2\frac{\dif\Gamma^{X\rightarrow W}_{\nu}(y)}{\dif y}D_{e}^{\nu_{2\alpha}}\left(\tfrac{x}{y}\right)  \Bigg],
\end{multline}
with flavor indices $\alpha$ and $\beta$. Additionally, the non-zero probability for neutrinos in the final state for ``no-splitting'' leads to an extra contribution
\begin{equation}
   D_{\nu_{2\alpha}}^{\nu_{2\alpha}}\left(\tfrac{x}{y},\tfrac{M_X}{2},0\right) \simeq \xi~\delta\left(\tfrac{x}{y}-1\right) ,
\end{equation}
\begin{figure}[t]
\centering
\includegraphics[width=0.5\textwidth]{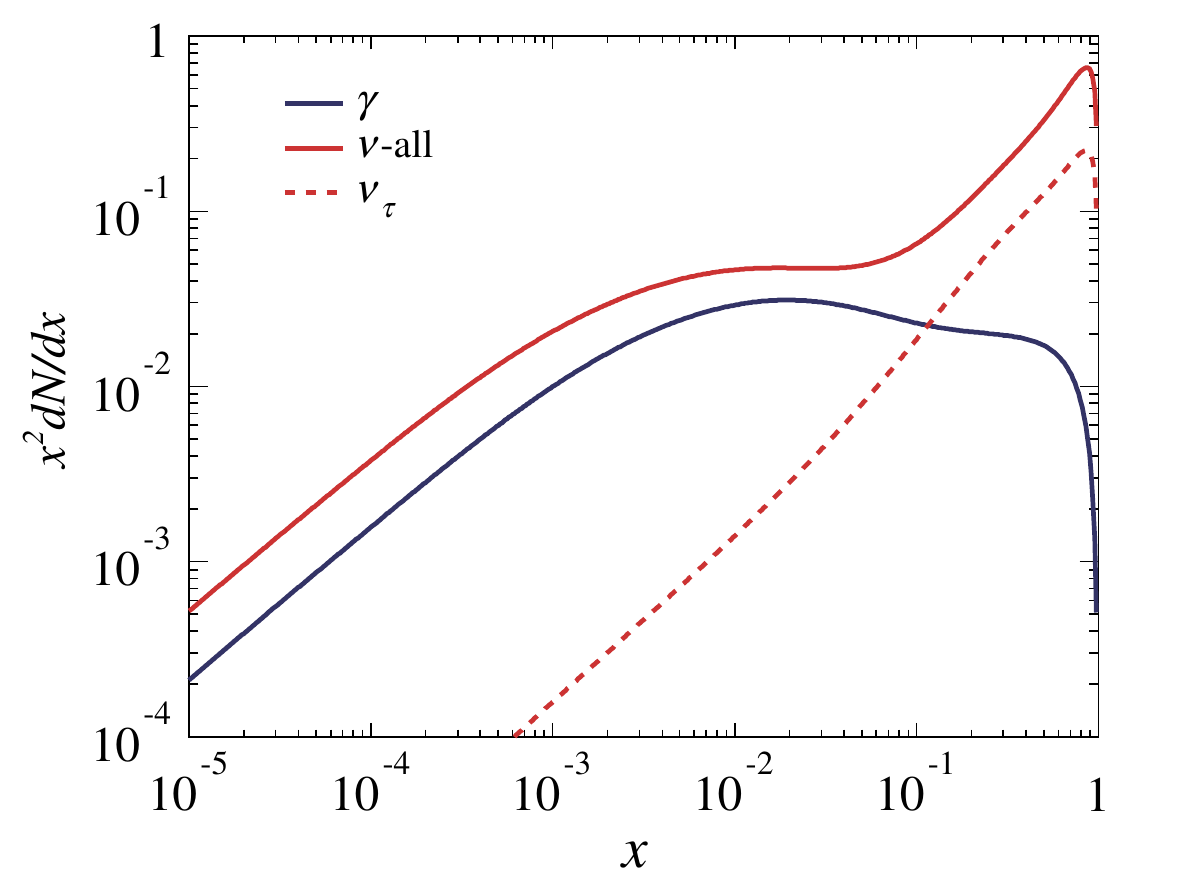}
\caption{Energy spectrum of neutrinos and photons from the decay of the pseudo-scalar particle $X$ within the BSM benchmark~\cite{Dudas:2020sbq} ($M_X=10^{10}~$GeV).}
\label{fig:energyspectra}
\end{figure}
with $\xi=0.13$. The contribution to the Earth-skimming detection mode is obtained similarly, considering only the $\tau$ flavor. Besides, the differential decay width into photons is also calculated following the same procedure, with proper FFs. Note that the expected photon fluxes to be compared to the flux limits are almost-entirely determined by the contribution of the Milky Way halo (due to their attenuation over inter-galactic scales). The resulting energy spectra are shown in 
Fig.~\ref{fig:energyspectra} for $M_X=10^{10}~$GeV. The high-energy enhancements are shaped by the non-zero probabilities for splitting a few times only at high scales. \\

\begin{figure}[htp]
\centering
\includegraphics[width=0.5\textwidth]{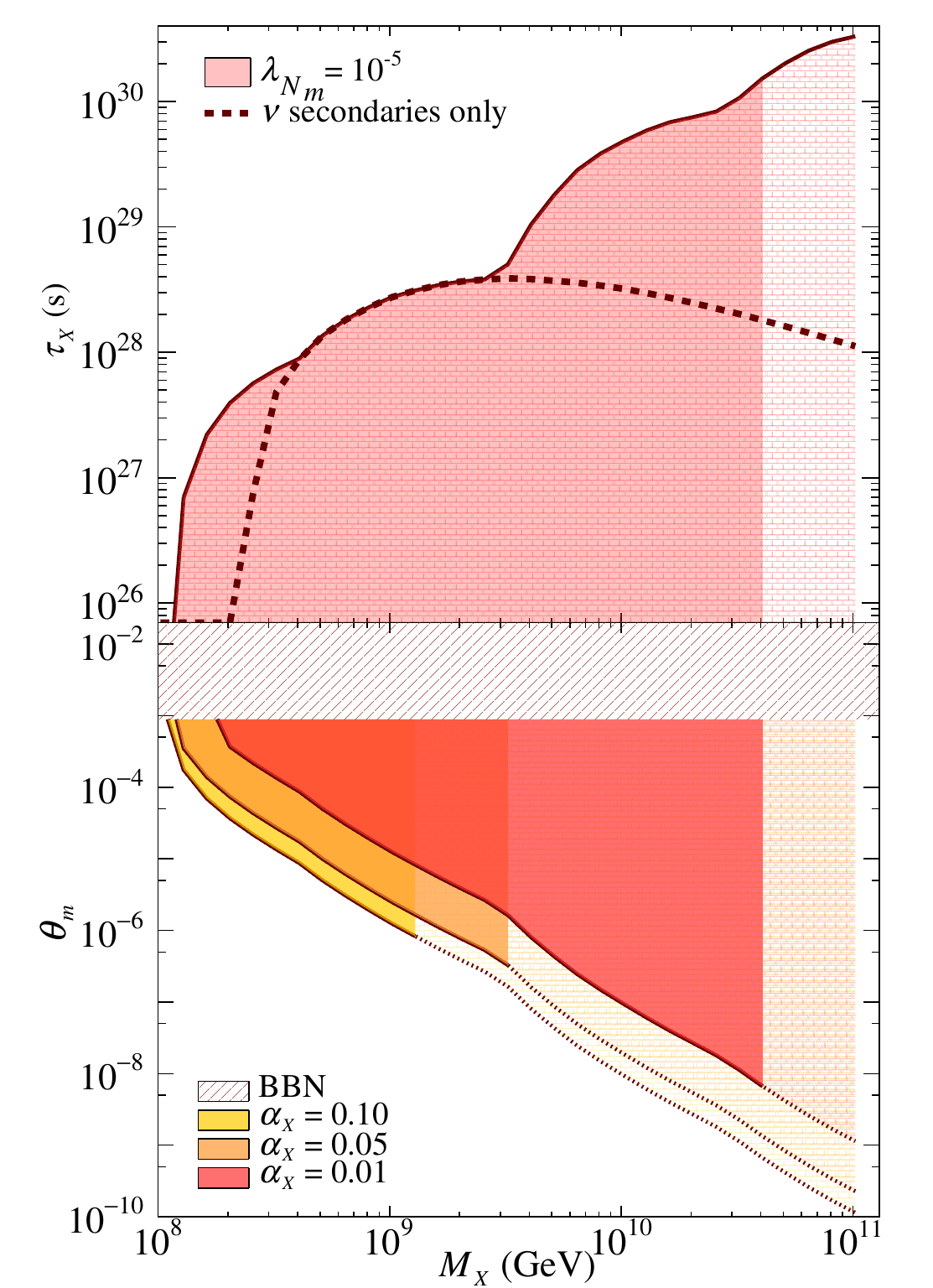}
\caption{\textit{Top}: Constraints on $\tau_X$ as a function of $M_X$ for a value of the couplings of sterile neutrinos with the inflationary sector $\lambda_{N_m}=10^{-5}$. The dotted line illustrates the constraints stemming from neutrino secondaries alone. \textit{Bottom}: Constraints on $\theta_m$ as a function of $M_X$ for three different values of the coupling constant $\alpha_{X}$, still for $\lambda_{N_m}=10^{-5}$. The hatched-red region $\theta_m\geq 9{\times}10^{-4}$ is excluded from the constraint on $\Delta N_\mathrm{eff}$ (see text).}
\label{fig:result}
\end{figure}

\textit{Constraints in the planes $(\tau_X,M_X)$ and $(\theta_m,M_X)$} can be derived from the non-observation of neutrinos at the Observatory. First, 90\% C.L. lower limits on the lifetime $\tau_X$ are obtained by setting, for a specific value of $M_X$, $n_\nu(E)$ or $n_\gamma(E)$ to the 90\% C.L. upper-limit numbers corresponding to the number of background-event candidates in the absence of signal~\cite{Feldman:1997qc,Conrad:2002kn}. More details about the upper-limit numbers for each specific analysis searching for neutrinos or photons are provided in the supplemental material. Subsequently, a scan in $M_X$ is carried out. It leads to a curve in the plane $(\tau_{X},M_X)$ that pertains to the energy threshold $E$ considered. By repeating the procedure for several thresholds, a set of curves is obtained, reflecting the sensitivity of a specific energy threshold to some range of mass $M_X$. The union of the excluded regions finally provides the constraints in the $(\tau_{X},M_X)$ plane. Results are shown in Fig.~\ref{fig:result} (top panel); lifetimes within the cross-hatched region are excluded. The region in full red pertains to a particular value of a Yukawa coupling $\lambda_{N_m}=10^{-5}$, the meaning of which will be explained below. To illustrate the contribution from each secondary at our disposal, we show as the dotted the contribution to the constraints stemming from neutrinos alone; an analysis of the IceCube exposure dedicated to the benchmark-scenario decay channels would likely provide better sensitivity for exploring masses $M_X\lesssim 10^{8.5}~$GeV. The lower limit on $\tau_X$ is then transformed into an upper limit on $\theta_m$ using the expressions of the total width of the particle $X$. Results are shown in the bottom panel for separate values of $\alpha_X$: color-coded regions pertain to $\lambda_{N_m}=10^{-5}$ while their extension (in cross-hatched) would require smaller values of  $\lambda_{N_m}$. Systematic uncertainties on $\theta_m$ constraints amount overall to $\simeq \pm 15\%$; they are dominated by those on the neutrino exposure~\cite{PierreAuger:2019ens}. The restricted ranges of $M_X$ for different $\alpha_X$ values come from the requirements not to overclose the universe with DM, while the exclusion hatched band comes from not altering the expansion history of the universe with the presence of ultra-light species such as sterile neutrinos $N_m$. We now briefly explain how these constraints are obtained.
\\

In addition to its couplings to the DM sector and to the SM one through the Higgs isodoublet, the sterile neutrino $N_m$ is also coupled to an inflationary sector in the BSM benchmark~\cite{Dudas:2020sbq}. This coupling, governed by a unique Yukawa parameter $\lambda_{N_m}$ for every $\nu_1$ neutrinos, yields to a ``radiative'' production of $X$ via a diagram similar to that depicted in Eqn.~\ref{eqn:3body-Xhnunu} (substituting $X$ by the inflaton $\Phi$ in the initial state, and $h$ and $\nu_2$ by $X$ and $\nu_1$ in the final states). Such a mechanism leads to a direct production of DM during the reheating period that can be sufficient, in general, to match the right amount of DM observed today~\cite{Kaneta:2019zgw}. In the BSM benchmark~\cite{Dudas:2020sbq}, values for $\lambda_{N_m}$ are then required to range preferentially around $10^{-5}$.  To infer the DM density $n_X$ produced mainly during the reheating epoch, we also consider the minimal setup of gravitational production of $X$ particles through the annihilation of SM (inflaton) particles as in~\cite{Garny:2015sjg} (as in~\cite{Mambrini:2021zpp}). In these conditions, $X$ particles can be produced as long as the collision rate of particles is larger than the expansion rate $H$ and/or as long as the inflaton field oscillates. By contrast, $n_X$ is prohibitively low to allow any thermal equilibrium for DM. The collision term in the Boltzmann equation is then approximated as a source term only. Overall, the Boltzmann equation reads as
\begin{eqnarray}
   \label{eqn:boltzmann}
   \frac{\dif n_X(t)}{\dif t}+3H(t)n_X(t)=\sum_i \overline{n}^2_i(t)\gamma_i+\overline{n}_\phi(t)\Gamma_{X\nu_1\nu_1}.
\end{eqnarray}
Here, the sum on the right hand side represents the contributions from the SM and inflationary sectors. Using, on the one hand, the evolution of the SM-matter and inflaton densities derived in~\cite{Chung:1998zb} and~\cite{Mambrini:2021zpp} respectively, and, on the other hand, the $\mathrm{SM}+\mathrm{SM} \rightarrow X+X$ and $\Phi+\Phi \rightarrow X+X$ reaction rates $\gamma_i$ derived in~\cite{Garny:2017kha} and~\cite{Mambrini:2021zpp} respectively, the present-day relic abundance of DM can then be related, using Eqn.~\ref{eqn:boltzmann} in the same way as in~\cite{PierreAugerCollaboration:2022tlw}, to the mass $M_X$ , the Hubble rate at the end of inflation $\Hinf$, and the reheating efficiency $\epsilon$ quantifying the duration of the reheating period ($\epsilon=1$ for an instantaneous reheating)~\cite{Garny:2015sjg}. As a result, viable couples of values for $(H_\text{inf},M_X)$ scale as $\Hinf \propto M_X^2$ up to a maximum value for $M_X$, which depend on $\epsilon$ and $\alpha_X$. This scaling is a consequence of the domination of the radiative-production process over the gravitational one as long as the allowed values of $H_\text{inf}$ are too small for a given $M_X$ value to generate significant particle production by gravitational interactions. For larger masses, the contribution from gravitational interactions added to the radiative production of $X$ leads to an overproduction of DM that overcloses the universe, and there is thus no longer solution. This explains why the color-coded regions extend up to some maximum values of $M_X$ in Fig.~\ref{fig:result}, for a benchmark value of $\lambda_{N_m}=10^{-5}$. To the right of the regions shown in cross-hatched, $\lambda_{N_m}$ would need to be smaller.

Another constraint on $\theta_m$ stems from the upper bound on the departure from 3 of the effective number of neutrino degrees of freedom $N_\mathrm{eff}$, $\Delta N_\mathrm{eff}=N_\mathrm{eff}-3$, inferred from cosmological observations. From Friedmann equation, the Hubble parameter is governed by the energy-density content, which, during the radiation-dominated epoch, receives contributions from all relativistic species such as photons, active neutrinos and possible sterile ones. A significant value of $\Delta N_\mathrm{eff}$ would thus change the dynamics of the expansion during the radiation era and thus impact on the cosmological microwave background  measurements as well as on the big bang nucleosynthesis. Therefore, the 95\%-confidence-level current limits on $\Delta N_\mathrm{eff}<0.3$~\cite{Planck:2018nkj} provide constraints on the temperature $T_{\nu_1}$ of $\nu_1$ neutrinos: 
\begin{equation}
   \Delta N_\mathrm{eff}=\frac{3}{2}\left(\frac{T_{\nu_1}}{T_{\nu_2}}\right)^4<0.3.
   \label{eqn:Tnu-dNeff}
\end{equation}
This relationship is valid at any time before the radiation era started, in particular at the time each neutrino species decoupled from the thermal bath when the expansion rate became larger than the interaction rate $R_{\nu_i}$ of the species $\nu_i$. Given that the ratio $H/R_{\nu_i}$ scales as $(T/T_{\star\nu_i})^3$~\cite{Weinberg:2008zzc} (denoting as $T_{\star\nu_i}$ the decoupling temperature of $\nu_i$), and given that the neutral-current-interaction rate of the $\nu_1$ species is that of $\nu_2$ but reduced by $\theta_m^2$, the relation $T_{\star\nu_1}=\theta_m^{-\nicefrac{2}{3}}T_{\star\nu_2}$ holds. As soon as $T$ dropped below $T_{\star\nu_1}$, the temperature of the freely expanding $\nu_1$ species fell as $T_{\nu_1}=(a_{\star 1}/a)T_{\star\nu_1}$ (with $a_{\star 1}$ the scale factor at the time of the decoupling of the $\nu_1$ species). At temperatures $T<T_{\star\nu_1}$, other particles that are still in thermal equilibrium, such as electrons and positrons, annihilate once $T$ becomes smaller than their mass and thus heat the rest of the relativistic species (photons) relative to the neutrinos. Entropy conservation in this ``freeze-out'' then gives rise to the relationship
\begin{equation}
   \left(\frac{T_{\nu_1}(T_{\star\nu_2})}{T_{\star\nu_2}}\right)^3=\frac{\mathcal{N}_{\star 2}}{\mathcal{N}_{\star 1}},
   \label{eqn:Tnustar}
\end{equation}
with $\mathcal{N}_{\star i}$ the number of degrees of freedom at $a=a_{\star i}$. Combining Eqn.~\ref{eqn:Tnu-dNeff} and Eqn.~\ref{eqn:Tnustar} with $\mathcal{N}_{\star 2}=43/4$ (accounting for two photon polarization states, three species of neutrinos and three of anti-neutrinos, and electrons and positrons each with two spin states) and $T_{\star\nu_2}=2$~MeV~\cite{Weinberg:2008zzc}, we obtain that $\mathcal{N}_{\star 1}\gtrsim 35.95$ and thus, using~\cite{Husdal:2016haj} to convert $\mathcal{N}$ into a temperature, the following bound on $\theta_m$,
\begin{equation}
   \theta_m\leq\left(\frac{T_{\star\nu_2}}{T_{\star\nu_1}}\right)^{\nicefrac{3}{2}}\simeq\left(\frac{2~\mathrm{MeV}}{215~\mathrm{MeV}}\right)^{\nicefrac{3}{2}}\simeq 9{\times}10^{-4}.
\end{equation}
This bound is shown as the cross-hatched band in Fig.~\ref{fig:result}.\footnote{Once translated in terms of effective sterile neutrino mass and $\Delta N_\mathrm{eff}$ (see e.g.~\cite{Bridle:2016isd}), the constraints on neutrino mass-squared difference between eigenstates and on mixing angles between mass and flavour eigenstates inferred from neutrino-oscillation experiments are within the dark red band.} \\

\textit{In conclusion,} we have shown that the data of the Pierre Auger Observatory provide stringent constraints on the angle $\theta_m$ mixing sub-eV sterile and active neutrinos in the context of an extension to the SM that couples the sterile neutrinos to a superheavy DM candidate~\cite{Dudas:2020sbq}. For a typical dark coupling constant of 0.1, the mixing angle $\theta_m$ must satisfy, roughly, $\theta_m \lesssim 1.5\times 10^{-6}(M_X/10^9~\mathrm{GeV})^{-2}$ for a mass $M_X$ of the dark-matter particle between $10^8$ and $10^{11}~$GeV. Future sensitivity to the effective number of neutrino degrees of freedom $\Delta N_\mathrm{eff}$ from cosmological observations will be complementary, as they will probe $\theta_m$ below $10^{-3}$ independent of the mass $M_X$. In particular, a value of $\Delta N_\mathrm{eff}$ departing significantly from 3 would call for sterile neutrinos and would fix the range of the mixing angle $\theta_m$. Our constraints would then be decisive in determining whether the new physics thus revealed in the neutrino sector is intimately related to that of superheavy DM. \\

\textit{Acknowledgments.} The successful installation, commissioning, and operation of the Pierre Auger Observatory would not have been possible without the strong commitment and effort from the technical and administrative staff in Malarg\"ue. We are very grateful to the following agencies and organizations for financial support:
Argentina -- Comisi\'on Nacional de Energ\'\i{}a At\'omica; Agencia Nacional de
Promoci\'on Cient\'\i{}fica y Tecnol\'ogica (ANPCyT); Consejo Nacional de
Investigaciones Cient\'\i{}ficas y T\'ecnicas (CONICET); Gobierno de la
Provincia de Mendoza; Municipalidad de Malarg\"ue; NDM Holdings and Valle
Las Le\~nas; in gratitude for their continuing cooperation over land
access; Australia -- the Australian Research Council; Belgium -- Fonds
de la Recherche Scientifique (FNRS); Research Foundation Flanders (FWO);
Brazil -- Conselho Nacional de Desenvolvimento Cient\'\i{}fico e Tecnol\'ogico
(CNPq); Financiadora de Estudos e Projetos (FINEP); Funda\c{c}\~ao de Amparo \`a
Pesquisa do Estado de Rio de Janeiro (FAPERJ); S\~ao Paulo Research
Foundation (FAPESP) Grants No.~2019/10151-2, No.~2010/07359-6 and
No.~1999/05404-3; Minist\'erio da Ci\^encia, Tecnologia, Inova\c{c}\~oes e
Comunica\c{c}\~oes (MCTIC); Czech Republic -- Grant No.~MSMT CR LTT18004,
LM2015038, LM2018102, CZ.02.1.01/0.0/0.0/16{\textunderscore}013/0001402,
CZ.02.1.01/0.0/0.0/18{\textunderscore}046/0016010 and
CZ.02.1.01/0.0/0.0/17{\textunderscore}049/0008422; France -- Centre de Calcul
IN2P3/CNRS; Centre National de la Recherche Scientifique (CNRS); Conseil
R\'egional Ile-de-France; D\'epartement Physique Nucl\'eaire et Corpusculaire
(PNC-IN2P3/CNRS); D\'epartement Sciences de l'Univers (SDU-INSU/CNRS);
Institut Lagrange de Paris (ILP) Grant No.~LABEX ANR-10-LABX-63 within
the Investissements d'Avenir Programme Grant No.~ANR-11-IDEX-0004-02;
Germany -- Bundesministerium f\"ur Bildung und Forschung (BMBF); Deutsche
Forschungsgemeinschaft (DFG); Finanzministerium Baden-W\"urttemberg;
Helmholtz Alliance for Astroparticle Physics (HAP);
Helmholtz-Gemeinschaft Deutscher Forschungszentren (HGF); Ministerium
f\"ur Innovation, Wissenschaft und Forschung des Landes
Nordrhein-Westfalen; Ministerium f\"ur Wissenschaft, Forschung und Kunst
des Landes Baden-W\"urttemberg; Italy -- Istituto Nazionale di Fisica
Nucleare (INFN); Istituto Nazionale di Astrofisica (INAF); Ministero
dell'Istruzione, dell'Universit\'a e della Ricerca (MIUR); CETEMPS Center
of Excellence; Ministero degli Affari Esteri (MAE); M\'exico -- Consejo
Nacional de Ciencia y Tecnolog\'\i{}a (CONACYT) No.~167733; Universidad
Nacional Aut\'onoma de M\'exico (UNAM); PAPIIT DGAPA-UNAM; The Netherlands
-- Ministry of Education, Culture and Science; Netherlands Organisation
for Scientific Research (NWO); Dutch national e-infrastructure with the
support of SURF Cooperative; Poland -- Ministry of Education and
Science, grant No.~DIR/WK/2018/11; National Science Centre, Grants
No.~2016/22/M/ST9/00198, 2016/23/B/ST9/01635, and 2020/39/B/ST9/01398;
Portugal -- Portuguese national funds and FEDER funds within Programa
Operacional Factores de Competitividade through Funda\c{c}\~ao para a Ci\^encia
e a Tecnologia (COMPETE); Romania -- Ministry of Research, Innovation
and Digitization, CNCS/CCCDI -- UEFISCDI, projects PN19150201/16N/2019,
PN1906010, TE128 and PED289, within PNCDI III; Slovenia -- Slovenian
Research Agency, grants P1-0031, P1-0385, I0-0033, N1-0111; Spain --
Ministerio de Econom\'\i{}a, Industria y Competitividad (FPA2017-85114-P and
PID2019-104676GB-C32), Xunta de Galicia (ED431C 2017/07), Junta de
Andaluc\'\i{}a (SOMM17/6104/UGR, P18-FR-4314) Feder Funds, RENATA Red
Nacional Tem\'atica de Astropart\'\i{}culas (FPA2015-68783-REDT) and Mar\'\i{}a de
Maeztu Unit of Excellence (MDM-2016-0692); USA -- Department of Energy,
Contracts No.~DE-AC02-07CH11359, No.~DE-FR02-04ER41300,
No.~DE-FG02-99ER41107 and No.~DE-SC0011689; National Science Foundation,
Grant No.~0450696; The Grainger Foundation; Marie Curie-IRSES/EPLANET;
European Particle Physics Latin American Network; and UNESCO.

We acknowledge for this work the support of the Institut Pascal at Universit\'e Paris-Saclay during the Paris-Saclay Astroparticle Symposium 2022, with the support of the P2IO Laboratory of Excellence (program “Investissements d’avenir” ANR-11-IDEX-0003-01 Paris-Saclay and ANR-10-LABX-0038), the P2I axis of the Graduate School of Physics of Université Paris-Saclay, as well as IJCLab, CEA, APPEC, IAS, OSUPS, and the IN2P3 master projet UCMN. Finally, we thank S. Cl\'ery, Y. Mambrini and M. Pierre for numerous discussions about the BSM benchmark studied in this work.

\bibliographystyle{apsrev4-2}
\bibliography{biblio}

\section*{Supplemental material}

We provide in this supplemental material a few summary plots and data related to the sensitivity of the Observatory to neutrinos and photons. All needed data can be found on www.auger.org.

\begin{figure}[htp]
\centering
\includegraphics[width=0.5\textwidth]{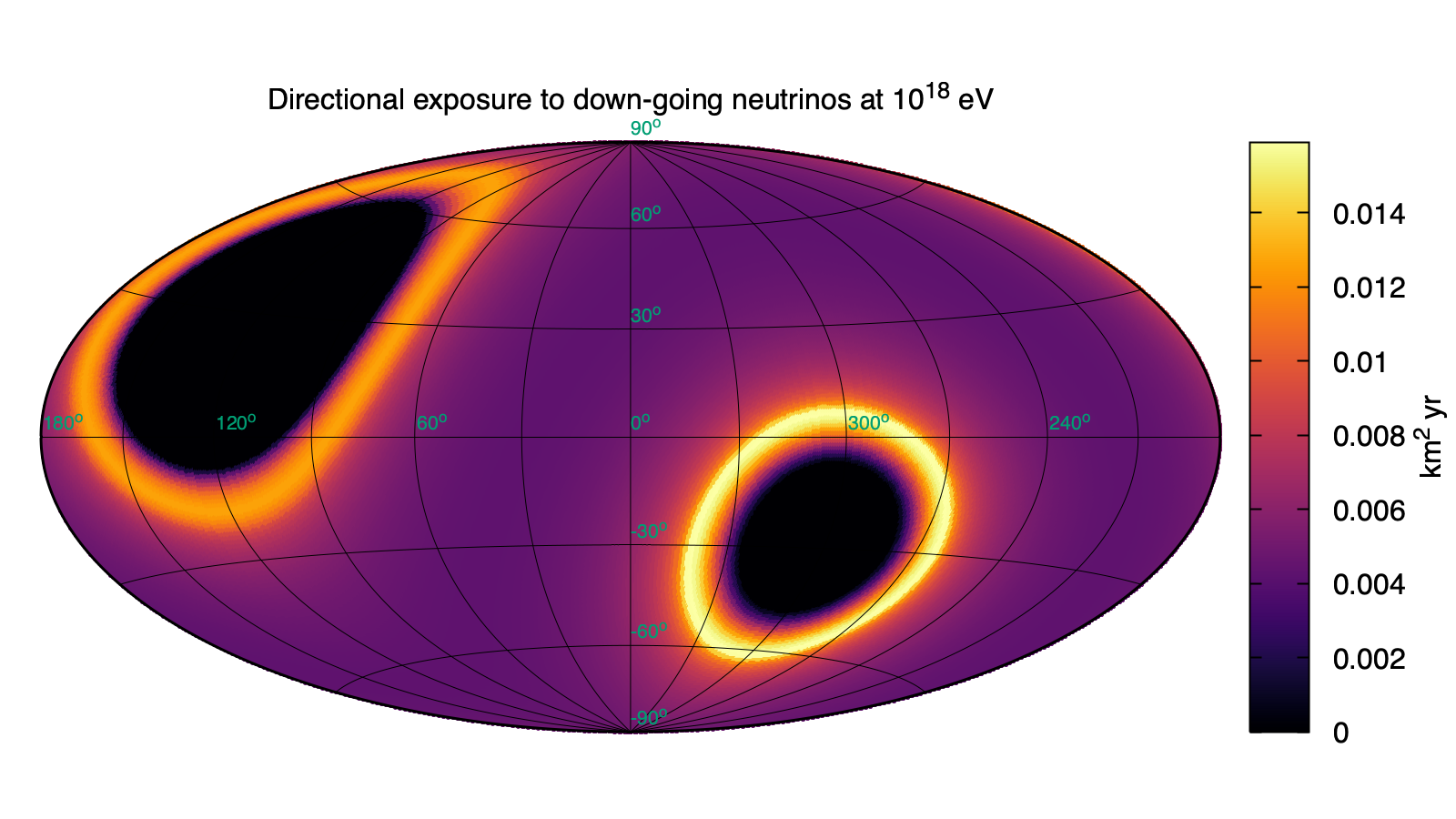}
\includegraphics[width=0.5\textwidth]{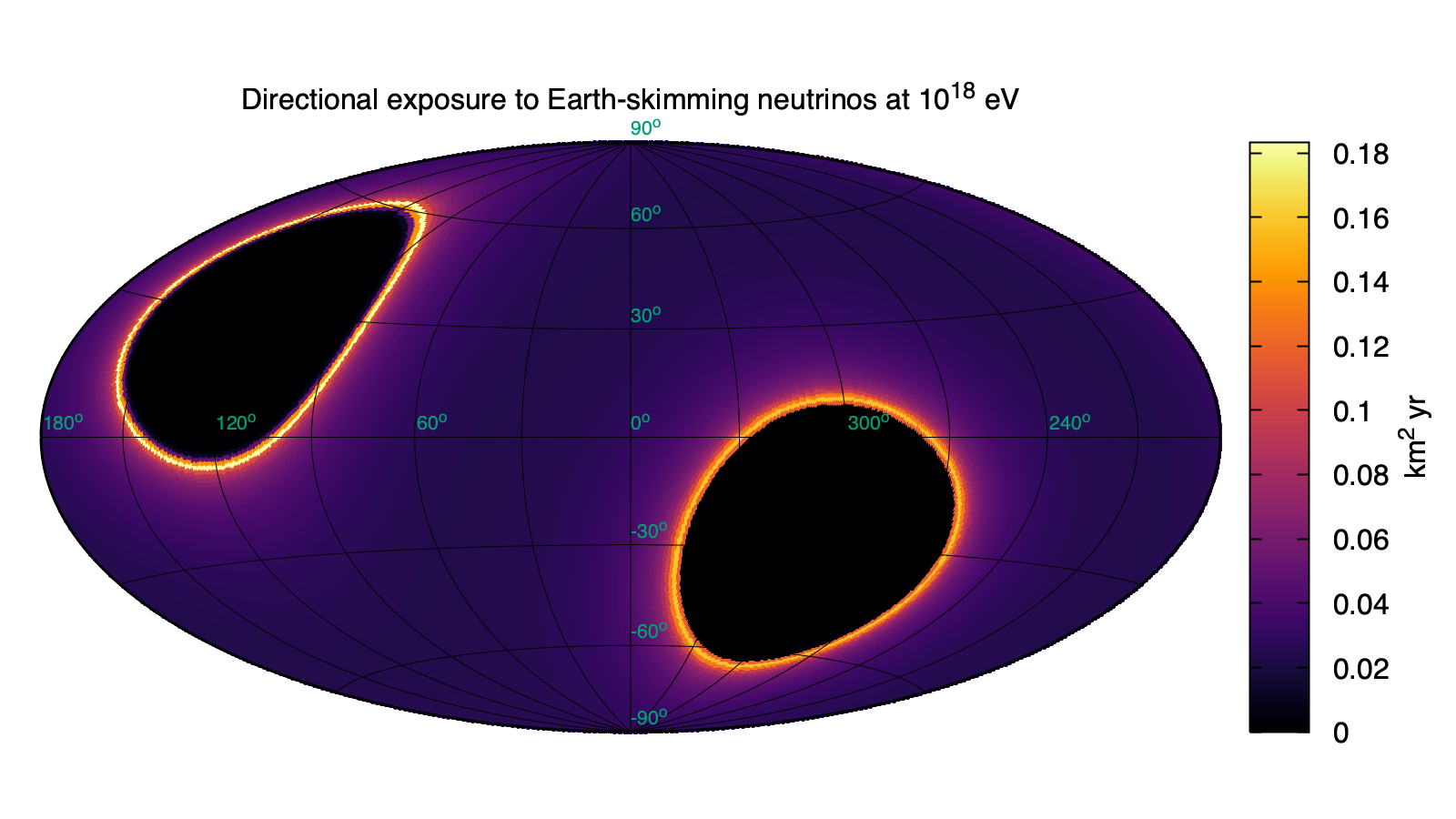}
\caption{Directional exposure of the Pierre Auger Observatory in Galactic coordinates to neutrinos at $10^{18}~$eV. \textit{Top}: Down-going channel, sensitive to all flavors. \textit{Bottom}: Earth-skimming channel, sensitive to $\nu_\tau$ flavor only.}
\label{fig:dir_expo_nu}
\end{figure}

The calculation of the exposure to neutrinos involves folding the detection efficiencies with the area of the 3,000~km$^2$ ground array projected onto the direction perpendicular to the arrival direction of the neutrino. In the down-going detection mode, the directional exposure in equatorial coordinates reads as
\begin{multline}
    \label{eqn:exponu-dg}
    \mathcal{E}^{\mathrm{DG}}_\nu(E,\delta)=\iiint \dif X\dif \mathbf{x}\dif t\cos{\theta}\epsilon_{i,c}(E,\theta,\varphi,X,\mathbf{x})\sigma_c m_\mathrm{p}^{-1},
\end{multline}
where the relationships relating local angles and time to equatorial angles are implicitly used, and where the integrations run over the depth at which the neutrino interaction can take place, the detector area and the time over the search period. The detection efficiency, $\epsilon_{i,c}$, depends on neutrino flavor $i$, the type of interaction $c$ (charged current or neutral current), neutrino energy $E$, zenith $\theta$ and azimuth $\varphi$ angles, the point of impact of the shower core on the ground $\mathbf{x}$, and the depth in the atmosphere $X$ measured along the shower axis at which the neutrino is forced to interact in the simulations. The term $\sigma_c m_\mathrm{p}^{-1} \dif X$, with $m_\mathrm{p}$ the mass of a proton and $\sigma_c$ the neutrino-nucleon cross-section, represents the probability of neutrino-nucleon interactions along a depth $\dif X$ expressed in g\,cm$^{-2}$. The result is illustrated in the top panel of Fig.~\ref{fig:dir_expo_nu} at $E=10^{18}~$eV in Galactic coordinates. The zero-exposure region in the Northern hemisphere reflects the absence of sensitivity to showers beyond 90$^\circ$ in zenith angle given the Earth latitude of $\simeq -39.25^\circ$ of the Observatory, while that in the Southern hemisphere reflects the minimum zenith angle of the analysis (75$^\circ$ here). In the bottom panel, the directional exposure is shown in the case of the Earth-skimming detection mode. In that case, the calculation is more involved. The efficiency $\epsilon_\mathrm{ES}$ depends on the energy of the emerging $\tau$ leptons $E_\tau$, the position of the signal pattern on the ground, the zenith and azimuth angles, and the altitude $h_\mathrm{dec}$ of the decay point of the $\tau$ above ground. $\epsilon_\mathrm{ES}$ is averaged over the decay channels of the $\tau$ with their corresponding branching ratios and is subsequently folded with the projected area of the detector, with the differential probability $p_\mathrm{exit}(E,\theta,E_\tau) = \dif p_\mathrm{exit}/\dif E_\tau$ of a $\tau$ emerging from the Earth with energy $E_\tau$, as well as with the differential probability $p_\mathrm{dec}(E_\tau,\theta,h_\mathrm{dec}) = \dif p_\mathrm{dec}/\dif h_\mathrm{dec}$ that the $\tau$ decays at an altitude $h_\mathrm{dec}$. The directional exposure is then 
\begin{multline}
    \label{eqn:exponu-es}
    \mathcal{E}^{\mathrm{ES}}_{\nu_\tau}(E,\delta)=\iiiint \dif \mathbf{x}\dif E_\tau\dif h_\mathrm{dec}\dif t|\cos{\theta}|\epsilon_\mathrm{ES}p_\mathrm{exit}p_\mathrm{dec}.
\end{multline}
The sky map shown in Fig.~\ref{fig:dir_expo_nu} (bottom panel) reflects that in local coordinates, the integrand of Eqn.~\ref{eqn:exponu-es} is non-zero in a narrow zenithal range of a few degrees only just below the horizon. \\

\begin{table}[h]
\caption{Upper limits on the number of photon candidates for each energy threshold.}
\label{tab:n}
\begin{ruledtabular}
\begin{tabular}{c c}
$E (10^{18}~\mathrm{eV})$ & $n_\gamma(E)$ \\
\colrule
$0.05$ & $2.44$  \\
$0.08$ & $2.44$  \\
$0.12$ & $2.44$  \\
$0.20$ & $2.44$  \\
$0.30$ & $2.44$  \\
$0.50$ & $2.44$  \\
$1.00$ & $23.38$  \\
$2.00$ & $9.53$  \\
$3.00$ & $3.43$  \\
$5.00$ & $2.59$  \\
$10.0$ & $24.00$  \\
$20.0$ & $4.36$  \\
$40.0$ & $2.44$  \\
\end{tabular}
\end{ruledtabular}
\end{table}

\begin{figure}[b]
\centering
\includegraphics[width=0.5\textwidth]{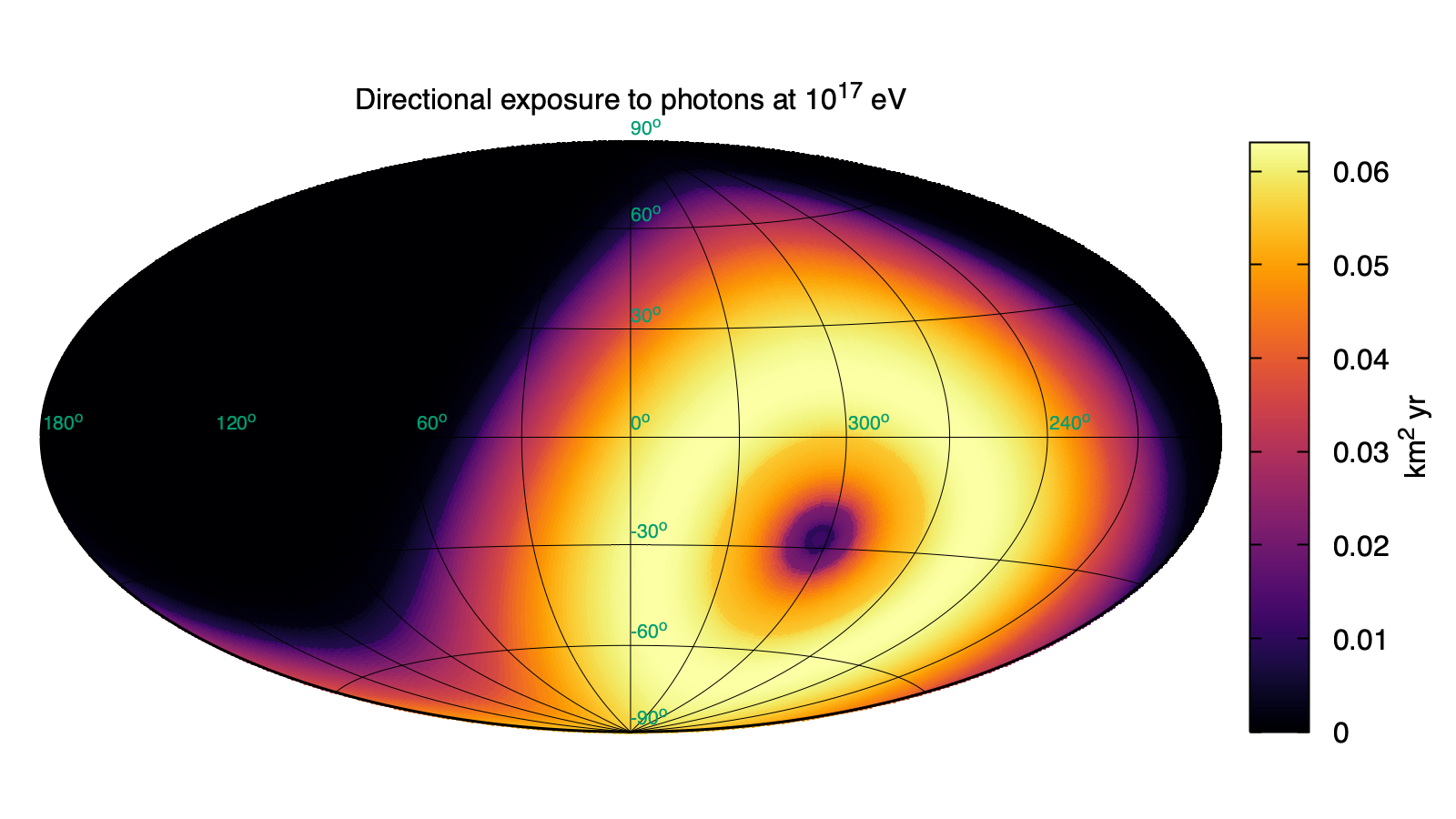}
\includegraphics[width=0.5\textwidth]{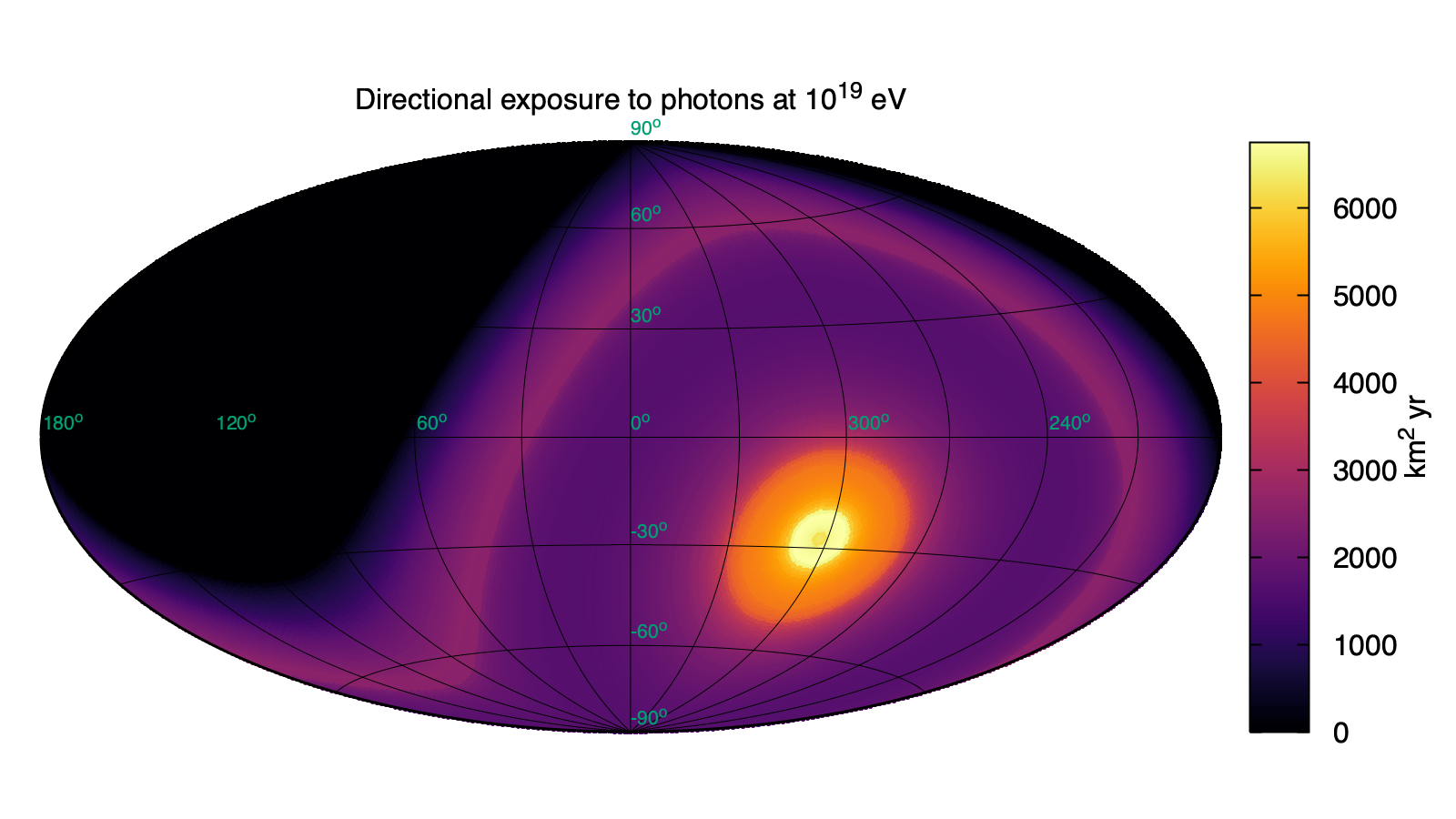}
\caption{Directional exposure of the Pierre Auger Observatory in Galactic coordinates to photons at $10^{17}~$eV (\textit{top}) and $10^{19.5}~$eV (\textit{bottom}).}
\label{fig:dir_expo_ph}
\end{figure}

In the case of photons, the directional exposure is obtained through the time and area integration of the detection efficiency $\epsilon_\gamma$ and selection efficiency $\kappa_\gamma$ projected onto the direction perpendicular to the arrival
direction of the photon,
\begin{equation}
    \label{eqn:expoph}
    \mathcal{E}_\gamma(E,\delta,\alpha)=\iint \dif \mathbf{x}\dif t \cos{\theta}\epsilon_\gamma(E,\theta,\mathbf{x},t)\kappa_\gamma(E,\theta).
\end{equation}
Four different analyses, differing in the detector used, have been developed to cover the wide energy range probed at the Observatory. In particular, two of these analyses benefit from a direct estimate of the depth of shower maximum by the fluorescence detector as one of the discriminating variables. The use of the fluorescence-detector datasets introduces for these analyses an explicit dependence in time for $\epsilon_\gamma$ due to the limited duty cycle on moonless nights that propagates into a small dependence in right ascension $\alpha$ for $\mathcal{E}_\gamma$. In addition to the detection efficiency, the $\kappa_\gamma$ factor accounts for the dependencies of the selection process aimed at separating hadronic-induced showers from photon-induced ones. By contrast to the neutrino case, searches for photons are not background-free and, for some of the energy thresholds explored, there is an irreducible small number of charged cosmic-ray events that resemble photon events despite the selection process. The values used for $n_\gamma(E)$ result from the upper limits on the number of candidates accounting for this irreducible background; they are listed in Table~\ref{tab:n}. For illustration, we show in Fig.~\ref{fig:dir_expo_ph} an example of directional exposure to photons at $10^{17}~$eV (top panel) and at $10^{19}~$eV (bottom panel). The former is built on the 1.14~km$^2$ array with a separation of detectors of 433~m optimized to study the range of energies around $10^{17}~$eV, while the latter on the 3,000~km$^2$ one, optimized for higher energies. This explains the large increase of exposure observed at high energies.

\end{document}